\documentclass[a4paper,10pt]{article}
\pdfoutput=1 

\usepackage{jheppub} 
\usepackage{multirow}

\usepackage[T1]{fontenc} 
\usepackage{hyperref}
\usepackage{graphicx}

\newcommand{\tb}{t_\beta}

\newcommand{\cba}{c_{\beta\alpha}}
\newcommand{\sba}{s_{\beta\alpha}}
\newcommand{\maa}{m_{A^0}}

\newcommand{\mhh}{m_{H^0}}
\newcommand{\mhp}{m_{H^\pm}}

\newcommand{\tti}{t\bar{t}}
\newcommand{\bb}{b\bar{b}}
\newcommand{\cc}{c\bar{c}}
\newcommand{\epem}{e^+e^-\rightarrow}

\newcommand{\fb}{\text{ fb}}
\newcommand{\iab}{\text{ ab}^{-1 }}
\newcommand{\ab}{\text{ ab}}
\newcommand{\gev}{\text{ GeV}}
\newcommand{\tev}{\text{ TeV}}
\newcommand{\myeq}{\stackrel{{\tiny{\cba\rightarrow 0}}}{=}}

\usepackage[pagewise]{lineno}

\title{Measuring the triple Higgs self-couplings in two Higgs doublet model}

\author{Nasuf SONMEZ}
\emailAdd{nasuf.sonmez@ege.edu.tr}
\affiliation{Department of Physics, Faculty of Science, Ege University, 35040, Izmir, Turkey}

\abstract{
    The determination of the Higgs self-coupling in the Standard Model is one of the primary motivations among all the future lepton colliders. Extending the scalar sector of the Standard Model by a new Higgs doublet with a quadratic Higgs potential gives many new features to the model, and most importantly additional Higgs self-couplings emerge. Measuring these couplings is the only way to reconstruct the shape of the scalar potential. In this study, the numerical analysis of several scattering processes is carried out for the two-Higgs-doublet model to determine all these Higgs self-couplings. These processes are selected among various possible combinations of additional Higgs states. The computation is carried out in the exact alignment limit ($\sba=1$). The distribution of the cross sections is presented regarding the polarization of the incoming beams and up to $\sqrt{s}=3\tev$. A strategy for extracting the Higgs self-couplings are considered in 2HDM and at the future lepton colliders. Possible final states that could be used for each of the processes are investigated using the decays of the final state particles.  
}

\begin{document}

\maketitle



\section{Introduction}
\label{sec1}

  The masses of the fundamental particles are generated through the electroweak symmetry breaking mechanism (EWSB) in the Standard Model (SM). That mechanism is constructed with the inclusion of a scalar Higgs field and quadratic scalar potential. One prominent prediction of this theory is a scalar particle named the Higgs boson which had been searched for a long time, and finally, it was discovered at the LHC \cite{Aad:2012tfa, Chatrchyan:2012xdj, Khachatryan:2014jba, Aad:2015gba}. Many production and decay channels of the Higgs particle have been studied extensively since then. According to the measurements, it resembles what the SM predicts so far. There is a model called the two-Higgs-doublet model (2HDM) where another Higgs doublet is added to the SM. In 2HDM, there are in total of five Higgs states: two charged ones $(H^\pm)$ and three neutral bosons $(h^0, A^0, H^0)$. Adding this extra scalar doublet introduces rich phenomenological implications which need attention.

    The experiments at the LHC goes well so far, and it produced many results. However, a precision machine such as lepton collider is necessary for studying the Higgs particle and its properties thoroughly. In lepton colliders, the initial state is well defined. If the four-momenta of the remnants in the collision could be extracted in high precision, that would help to reconstruct the event in detail. There are couple of proposals for the future lepton colliders: the Circular Electron-Positron Collider (CEPC) in China \cite{Gao:2017ubc, Liang:2016mue, Xiao:2015vrz}, the Future Circular Collider (FCC-ee) \cite{fcc-ee} at CERN \cite{Gomez-Ceballos:2013zzn}, and the International Linear Collider (ILC) in Japan \cite{Yamamoto:2017lnu}. These proposals support that the electron-positron colliders are the excellent choice to produce many Higgs bosons and to study its properties. They will be literally the Higgs factories, and they could be used for complementing all the LHC results.

    According to the SM, the mass of the Higgs particle is related to Higgs self-coupling by $m^2_H = 2\lambda v^2$ at the tree level, and measuring the Higgs mass makes it possible to determine the Higgs self-coupling $g_{HHH}$. To establish the EWSB mechanism thoroughly, the scalar potential of the Higgs field needs to be constructed orthogonally. That requires making measurements of the triple and the quadratic self-couplings, $g_{HHH}$ and $g_{HHHH}$ respectively. Studying the double Higgs-strahlung ($e^-e^+ \rightarrow ZHH$),  along with WW double-Higgs fusion ($e^-e^+ \rightarrow \nu\bar{\nu}HH$) makes it possible to measure the triple Higgs self-coupling with astonishing precision in the SM \cite{GutierrezRodriguez:2008nk, Battaglia:2001nn, dEnterria:2016sca, Castanier:2001sf}. On the other hand, measuring Higgs self-coupling allows us to reconstruct the Higgs potential in the SM, which is the most conclusive test of the EWSB mechanism. If the scalar sector is extended like the 2HDM, determining the self-couplings, as well as the Higgs potential, could be a complicated task. In 2HDM, there are in a total of 8 trilinear Higgs self-couplings. A similar attempt was made before at what extent the trilinear Higgs couplings could be probed by studying various Higgs boson pairs associated with the Z boson in reference \cite{Arhrib:2008jp}. However, the processes and the region of interest differs from this study, and most importantly the motivation for the free parameters of the model does not hold the primary theoretical constraints (perturbativity and unitarity) of the model which was claimed otherwise. Some of the couplings were studied through the double and the triple Higgs boson production in references \cite{Ferrera:2008nu, Ferrera:2007sp}. Besides, triple and quartic Higgs couplings have been studied at the linear colliders in the context of the MSSM in references \cite{Djouadi:1999gv, Dubinin:1998nt, Dubinin:2002nx, Chalons:2017wnz, Boudjema:2001ii, Muhlleitner:2000zu}. In this work, we analyzed various scattering processes in $e^-e^+$-collider and concluded whether all these Higgs self-couplings could be determined. In this aim, the correlation between these couplings and scattering processes is examined, and a plan is offered on how to determine the triple Higgs self-couplings. Distributions for the cross sections are calculated as a function of the center-of-mass (c.m.) energy and the polarization of the incoming beams. The results are obtained in 2HDM is analyzed for the free parameters of the model considering the recent experimental limits.

    This paper is organized as follows. In section \ref{sec2}, we briefly introduce the scalar potential, the relevant couplings and the Higgs mechanism in SM and 2HDM. A discussion on the experimental and theoretical constraints are carried out. The connection between the processes and the Higgs self-couplings are analyzed. In section \ref{sec3}, the analytical expressions regarding the kinematics of the scattering are presented. The numerical results and a discussion are given in section \ref{sec4}. In section \ref{sec5}, the decay products of the Higgses and the identification of each of the processes is examined. At last, the conclusion and summary are delivered in section \ref{sec6}.

\section{Short review of the Higgs mechanism and the self-couplings in 2HDM}
\label{sec2}

\subsection{Higgs mechanism in the SM}

    In the SM, the electroweak gauge bosons and the fundamental matter particles acquire their masses interacting with a scalar field called the Higgs field. The scalar potential is defined as follows:
    \begin{equation} \label{eq:eq1}
    V(\Phi)=\mu^2|\Phi|^2+\frac{1}{2}\lambda|\Phi|^4. 
    \end{equation}
    The Mexican shape like potential is obtained when the parameters $\mu$ and $\lambda$ have the opposite sign. To guarantee the stability of vacuum, the self-coupling $\lambda$ is assumed to be positive, and $\mu^2 < 0$ is set accordingly. The minimum of the scalar potential occurs at $\langle\Phi\rangle = v= 246\gev$ where $\mu^2 = -\lambda v^2$.
    If the Higgs field is expanded around its vacuum expectation value, and the scalar potential sorted out, then we get the mass term of the Higgs particle and the Higgs self-couplings \cite{Baer:2013cma, Novaes:1999yn}
        \begin{equation}
            m^2_H = 2\lambda v^2 ,\;\;\;\; g_{HHH} = -3i\lambda v, \;\;\;\; g_{HHHH} = -3i\lambda.
        \end{equation}
    The complete reconstruction of the Higgs potential in the SM requires the determination of the trilinear ($g_{HHH}$) and the quadratic ($g_{HHHH}$) Higgs self-couplings.

\subsection{Scalar potential and parameter space in the 2HDM}

    In this section, we give a summary of the scalar potential and the parameters which are relevant to the results. 2HDM simply includes a second $SU(2)_L$ Higgs doublet with the same hypercharge of the original Higgs field. This model has been studied extensively in the literature \cite{Branco:2011iw, Gunion:1989we, Haber:2006ue, Davidson:2005cw, Carena:2002es}. Since we are not interested in the flavor-changing-neutral-currents (FCNCs) in this study, a discrete symmetry called $\mathcal{Z}_2$ is imposed on the Lagrangian \cite{Glashow:1976nt} which constrains them. 
    The Higgs doublets in Higgs basis are defined as $\Phi_i,\;(i=1,2)$ where
    \begin{equation}
        \Phi_1 = 
        \begin{pmatrix}
        G^+  \\
        \frac{1}{\sqrt{2}}[ v + S_1 + iG^0]\\ 
        \end{pmatrix},    
        \;\;
        \Phi_2 = 
        \begin{pmatrix}
        H^+  \\
        \frac{1}{\sqrt{2}}[ S_2 + i S_3 ] \\ 
        \end{pmatrix}.
        \label{eq:eq2}
    \end{equation}

    Accordingly, the scalar potential in the Higgs basis is defined in equation \ref{eq:eq3}.
        \begin{eqnarray}        
        V(\Phi_1,\Phi_2) = &m_{1}^2 &| \Phi_1|^2+m_{2}^2|\Phi_2|^2 - \left[ m_{3}^2   \Phi_1^{\dagger} \Phi_2 +h.c. \right] \nonumber \\ 
        &+& \frac{\Lambda_1}{2}| ( \Phi_1^{\dagger} \Phi_1 )^2  + \frac{\Lambda_2}{2} ( \Phi_2^{\dagger} \Phi_2 )^2 +\Lambda_3 | \Phi_1 |^2 | \Phi_2 |^2 +\Lambda_4 | \Phi_1^{\dagger} \Phi_2 |^2 \label{eq:eq3}  \\
        &+& \left [  \frac{\Lambda_5}{2} (\Phi_1^{\dagger} \Phi_2)^2  + h.c. \right]
        +\left[ \left(\Lambda_6 \Phi_1^{\dagger} \Phi_1 + \Lambda_7\Phi_2^{\dagger} \Phi_2\right) \Phi_1^{\dagger} \Phi_2 + h.c.\right]   \nonumber 
        \end{eqnarray}
        where all the coupling constants are real. In general, the parameters $m_{3}^2$, $\Lambda_{5}, \Lambda_{6}$, and $\Lambda_{7}$ could be complex, but we take them real for simplicity. Following the prescriptions defined in references \cite{Kanemura:2015ska, Haber:1978jt, Branco:2005jr}, the masses of all the extra Higgs bosons could be calculated as usual. That is simply plugging in the Higgs doublets into the scalar potential, and after sorting out the terms in equation \ref{eq:eq3}, the potential will be decomposed into a quadratic term plus cubic and quartic ones. The quadratic terms define the physical Higgs states and their masses. The masses are obtained by diagonalizing the quadratic mass terms. The rotation angle $\sba=\sin(\beta-\alpha)$ defines the mixing among the CP-even Higgs states \cite{Davidson:2005cw}. The rest of cubic and quartic terms define the couplings and the interactions among the new states in 2HDM.

    In this study, we explored the exact alignment limit and set $\sba=1$, as a result, $h^0$ becomes indistinguishable from the SM Higgs boson regarding mass and couplings. Consequently, the free parameters of the model which are essential for this study are the masses of the neutral Higgs bosons $(m_{h/H^0/A^0})$, the ratio of the vacuum expectation values ($\tan\beta$), the mixing angle between the CP-even neutral Higgs states ($\sba$), and the soft breaking scale of the discrete symmetry $(m^2_3)$ \cite{Gunion:2002zf}. It should be noted that the $m_3^2$ term in equation \ref{eq:eq3} ensures the breaking of the discrete symmetry softly. 
    
    These free parameters need to be constrained in some way. We imposed the following constraints, which are solely defined on a theoretical point of view in 2HDM. 
    \begin{itemize}
        \item Stability : The scalar potential has to be positive at large values of the field \cite{el2007consistency, sher1989electroweak, deshpande1978pattern, Nie:1998yn, ElKaffas:2006gdt}. 
        \item Unitarity : The amplitudes need to be flat at asymptotically large energies \cite{ginzburg2005tree}. 
        \item Perturbativity : All the quartic scalar couplings in 2HDM need to be smaller than a particular value, $|C_{H_i H_j H_k H_l} |<8\pi$. 
    \end{itemize}
    The parameter space is tested whether they obey these constraints with the help of \textsc{2HDMC-v1.7.0)} \cite{Eriksson:2009ws}.

\begin{figure}[htbp]
    \centering 
    \includegraphics[width=.45\textwidth]{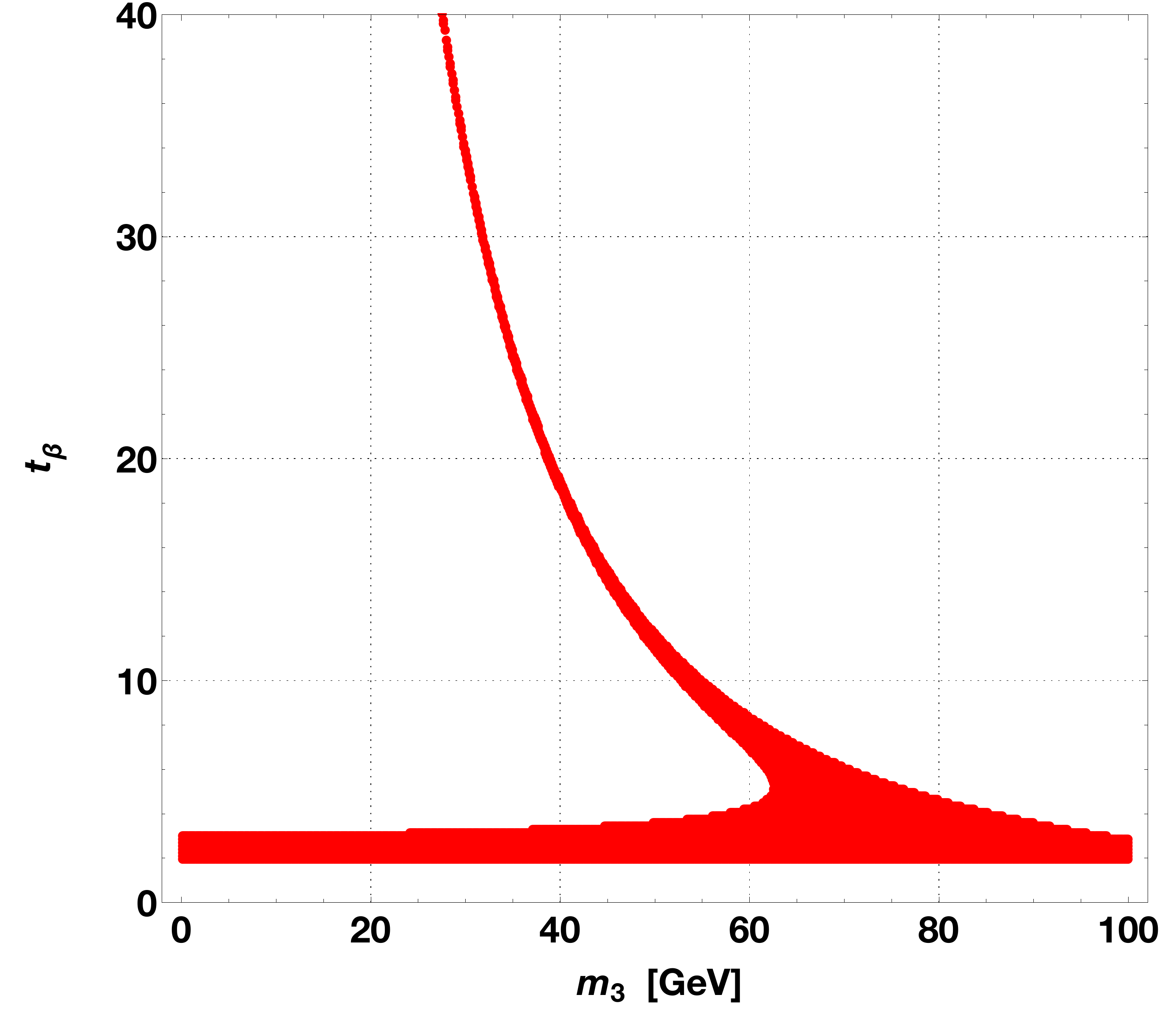}\\
    \caption{\label{fig6} The region of interest, where stability, unitarity, and perturbativity constraints are obeyed, is plotted as a function of the soft symmetry breaking term $m_{3}^2$ versus $\tb$.}
    \end{figure}

    There is another set of constraints which are coming from all the previous experiments. We followed a recent study \cite{Enomoto:2015wbn} where the flavor limits are presented, and figure 3 in reference \cite{Enomoto:2015wbn} particularly gives the available region which is not yet excluded. The 2HDM has charged Higgs states ($H^\pm$) compared to the SM, and these could easily make a novel contribution to the flavor observables. Besides, LEP, Tevatron, and LHC established many constraints on $\mhp$ and $t_\beta$. Discussion on the new limits is carried out in reference \cite{Moretti:2016qcc} and the references therein. Inspired by the reference \cite{Enomoto:2015wbn} and the current experimental results at the LHC \cite{Moretti:2016qcc}, masses of all the extra Higgs bosons are set to be $m_{H}=m_{H^0}= m_{A^0} = m_{H^\pm}$. This selection also minimizes the oblique parameters \cite{Peskin:1990zt, Peskin:1991sw, Haber:2010bw, gunion2003cp, Grimus:2008nb, Polonsky:2000rs}, so all the electroweak observables are close to the SM ones.  In the exact alignment limit, the decay of the $H^0$ to vector boson pairs is suppressed. On the other hand, as it is stated in reference \cite{Enomoto:2015wbn}, the neutral meson mixings $\Delta M_s$ in Type-I and the results of $\bar{\mathcal{B}}(B_s^0 \rightarrow \mu^+\mu^-)$ restrict the low $\tb$ region. Thus, $\tb>2$ region is adopted. Moreover, $m_{H}$ is not constrained for large $\tb$ range. As a result, the analysis is carried out in the $2<\tb<40$ range. Finally, the last parameter is the soft symmetry breaking term $m_{3}^2$ defined in equation \ref{eq:eq3}. The region, where the $m_{3}^2$ obeys the theoretical constraints, is given in figure \ref{fig6}, that region is calculated with the help of \textsc{2HDMC}, and accordingly, the central point is picked in the calculation. 
     
    The computation is performed in Type-I. On the other hand, the other types of Yukawa coupling schemes are also possible such as Type-I/-II/-Y/-X \cite{Branco:2011iw}. However, that does not affect the production rates of any of the process considered in this study because the couplings between the electrons and the positrons to Higgses does not make notable contributions. That will be explained more in the next section. The parameter region considered are presented in table \ref{tab1}.
    \begin{table}[htbp]
    \caption{The range of the free parameters of the model, all masses are given in GeV.\label{tab1}}     
    \centering
    \begin{tabular}{ c | ccccc }       
    \hline
    Benchmark    & Yuk. T.    & $m_{h^0}$  & $m_{(H^0/A^0/H^\pm)}$  & $\sba$  &    $\tb$   \\ \hline \hline        
    1            & Type-I     & 125        & (150..500)             & 1.0     &    (2..40) \\ 
    \hline \hline
    \end{tabular}
    \end{table}

\subsection{Higgs self-couplings in 2HDM}

    For completeness, we present the Higgs self-couplings in 2HDM as a function of the $\Lambda_i$ given in equation \ref{eq:eq3}. According to the parameter space, $\sba=1$, and $\cba=0$ are set. Besides, in the limit of $m_{H^0}= m_{A^0} = m_{H^\pm}$ and exact alignment the parameters $\Lambda_4$, $\Lambda_5$ and $\Lambda_6$ vanish. Therefore, the Higgs self-couplings get simplified, and they are given in equations \ref{eq:eq5} - \ref{eq:eq10}, where $\Lambda_{345}=\Lambda_3+\Lambda_4+\Lambda_5$. As the experimental results favor the exact alignment limit, it is illuminating whether the self Higgs couplings are possible to measure in this limit. Among all the Higgs self-couplings only $g_{h^0h^0H^0}$ vanishes, and the rest of them reduces down to a simple function of $\Lambda_i$.  Moreover, the couplings $g_{h^0H^0H^0}$ and $g_{h^0A^0A^0}$ are equal to each other, and we also remark that ratio is $g_{H^0H^0H^0}/g_{H^0A^0A^0}=3$. These predictions could also be tested experimentally. 

           \begin{eqnarray} 
            g_{h^0h^0h^0}&=&-3 i v ( (\Lambda_7 \cba^2+3 \Lambda_6 \sba^2)\cba+ (\Lambda_{345} \cba^2+\Lambda_{1} \sba^2)\sba)  \nonumber\\
                        &\myeq     &    -3iv\Lambda_1 \label{eq:eq5} \\
            g_{h^0h^0H^0}&=&-i v ( (\Lambda_{345} (1 - 3 \sba^2) + 3 \Lambda_1 \sba^2)\cba +  3 (\Lambda_6 (2 - 3 \sba^2) - \Lambda_7 \cba^2) \sba)\nonumber\\
                        &\myeq     &    0 \label{eq:eq6} \\
            g_{h^0H^0H^0}&=& -i v ((3 \Lambda_1 \cba^2 + \Lambda_{345} ( 3 \sba^2-2)) \sba + 3  (\Lambda_6+\Lambda_7 \sba^2-3 \Lambda_6 \sba^2)\cba ) \nonumber \\
                        &\myeq     &    -iv\Lambda_3 \label{eq:eq7} 
            \end{eqnarray}

            \begin{eqnarray} 
            g_{h^0A^0A^0}&=&-i v (\Lambda_7\cba+(\Lambda_3+\Lambda_4-\Lambda_5 )\sba)\nonumber\\
                        &\myeq     &    -iv\Lambda_3 \label{eq:eq8} \\
            g_{H^0H^0H^0}&=&-3 i v ( (\Lambda_{1} \cba^2+\Lambda_{345} \sba^2)\cba-\Lambda_7 \sba^2-3 \Lambda_6 \cba^2 )\sba\nonumber\\
                        &\myeq     &    3iv\Lambda_7 \label{eq:eq9} \\
            g_{H^0A^0A^0}&=&-i v ((\Lambda_3+\Lambda_4-\Lambda_5)\cba-\Lambda_7 \sba)\nonumber\\
                        &\myeq     &    iv\Lambda_7 \label{eq:eq10} 
            \end{eqnarray}
        

        \subsection{Determining the Higgs self-couplings in 2HDM}

\begin{figure}[htbp]
    \centering 
    \includegraphics[width=.65\textwidth]{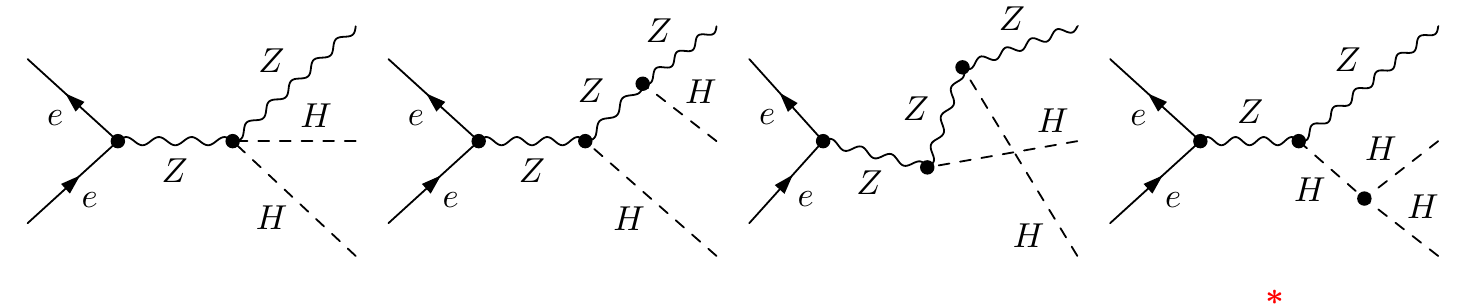}\\
    \caption{\label{fig1} The Feynman diagrams which contribute to the scattering process of $e^-e^+\rightarrow Z^0HH$ at the tree level in the SM. The amplitude with the red dot is the only one includes the coupling $g_{HHH}$ and also makes the dominant contribution.}
    \end{figure}

    In the SM, due to the small coupling between the Higgs boson and electron-positron $g_{e^-e^+H}$, diagrams where the Higgs boson is intermediated do not make a noticeable contribution. Therefore, they could be neglected safely. The contributing Feynman diagrams are given in figure \ref{fig1}. The diagram with the red star is the one which makes the dominant contribution to the production of $ZHH$, and that diagram alone makes it possible to measure the Higgs self-coupling in the SM. Moreover, the quartic coupling $g_{ZZHH}$ is also suppressed compared to the trilinear coupling $g_{HHH}$.
    
    The situation is cumbersome for the 2HDM because there is more than one Higgs self-coupling. As we are interested in the case where the $h^0$ is indistinguishable from the SM Higgs ($H$), the arguments in the SM hold for the 2HDM as well. The absolute value of the couplings $g_{e^-e^+h^0}$, $g_{e^-e^+H^0}$ and $g_{e^-e^+A^0}$ are less than $\sim10^{-6}$. Therefore, they could be neglected and noted that $g_{e^-e^+Z^0}$ coupling is the only one which could make a significant contribution. Thus, Feynman diagrams, where the Z-boson is intermediated, are the ones we take into account in the computation. At last, the quadratic couplings compared to the trilinear Higgs self-couplings are small, and they could be omitted as well. The first set of scattering processes which are investigated includes the following final states. They are $Z^0H^0h^0$, $Z^0A^0h^0$, $H^0H^0H^0$, $h^0h^0h^0$, $A^0h^0h^0$, $A^0A^0h^0$, $A^0A^0H^0$, $H^0h^0h^0$, and $H^0H^0h^0$. All these processes include various combination of the trilinear Higgs self-couplings, but in any case, their cross sections are less than $\sim10^{-11}\fb$. The only exception is the process $e^-e^+\rightarrow A^0A^0A^0$ which is at the order of $\sim 0.04 \ab$. However, it still less than an atto-barn, therefore, it could not be possible to detect a single event throughout the lifetime of the proposed colliders. Hence, the second set of processes are given in the top row of table \ref{tab:tab2}. These are the scattering processes which could be used to determine the Higgs self-couplings in the 2HDM. Moreover, they are the only ones which have a cross section greater than atto barn. In table \ref{tab:tab2}, the trilinear Higgs self-couplings, which contribute to the scattering process indicated on top of each column, are marked by a plus sign. The coupling $g_{h^0h^0H^0}$ vanishes as it is given in equation \ref{eq:eq6}. Measuring the cross section of $e^-e^+\rightarrow Z^0A^0A^0$ lets us to determine the coupling ${g_{h^0A^0A^0}}$. Next, studying the $e^-e^+\rightarrow Z^0H^0H^0$ makes it possible to determine ${g_{h^0H^0H^0}}$. The coupling $g_{h^0h^0h^0}$ (which is also in the SM) could be determined with the same scattering process $e^-e^+\rightarrow Z^0h^0h^0$. Accordingly, the coupling $g_{H^0A^0A^0}$ could be determined by studying $e^-e^+\rightarrow A^0H^0h^0$. Finally, $g_{H^0H^0H^0}$ could be extracted from $e^-e^+\rightarrow A^0H^0H^0$ with the coupling ($g_{H^0A^0A^0}$) obtained in the previous step.

    \begin{table}[htp]
    \caption{The trilinear Higgs self-couplings contributing to each scattering process in a future linear collider. The exact alignment $\sba=1$ and $m_{H}=\mhh=\maa$ are set. \label{tab:tab2}
    }
    \centering
    \begin{tabular}{ l || c | c | c | c | c }
    \hline  
                              & $ Z^0A^0A^0$ 
                                  & $ Z^0H^0H^0$
                                      & $ Z^0h^0h^0$    
                                          & $A^0H^0h^0$ 
                                               & $ A^0H^0H^0$
                                                \\
    \hline\hline                
    ${g_{h^0h^0h^0}}$    &   &   & + &  &       \\ \hline
    ${g_{h^0H^0H^0}}$    &   & + &   &   &      \\ \hline
    ${g_{H^0H^0H^0}}$    &   &   &   &   & +      \\ \hline
    ${g_{h^0A^0A^0}}$    & + &   &   &   &       \\ \hline
    ${g_{H^0A^0A^0}}$    &   &   &   & + & +     \\ 
    \hline 
    \end{tabular}
    \end{table}%

\begin{figure}[htbp]
    \centering 
    \includegraphics[width=.90\textwidth]{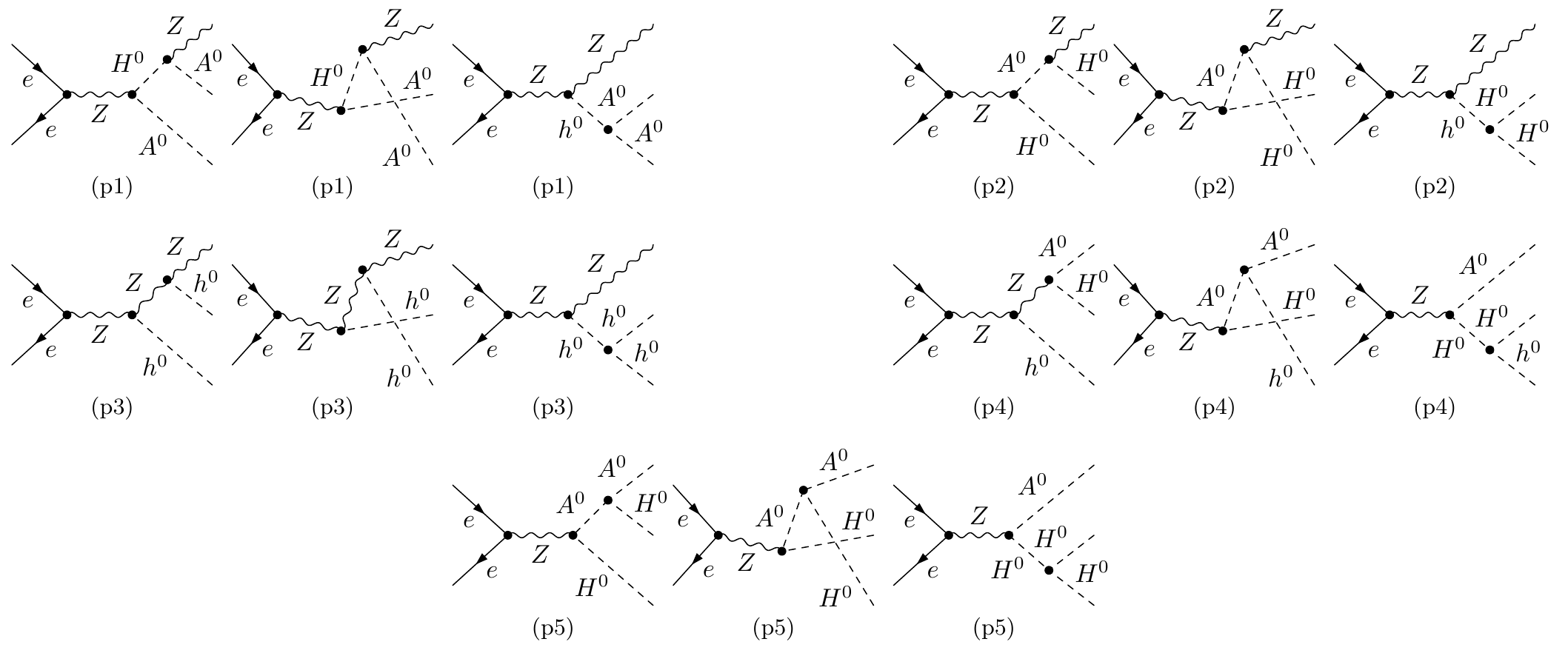}
    \caption{\label{fig2}
   The set of Feynman diagrams, which contribute to the scattering process given in table \ref{tab:tab2}, with the caption indicate which scattering \textbf{p}\emph{rocess} it belongs to. For example; p1 stands for the first process in the top row ($e^-e^+\rightarrow Z^0A^0A^0$ ) in table \ref{tab:tab2}. It should also be noted that the calculation is carried out in the Feynman gauge, and there are the same set of diagrams with a Goldstone boson ($G^0$) which is intermediated instead of a vector boson.}    \end{figure}

    The Feynman diagrams which contribute to each scattering process are given in figure \ref{fig2}. They all share the same topology, Z-boson is intermediated between the initial and the final states, but different particles and couplings are involved. Besides of all these self-couplings, it can be seen at first glance in figure \ref{fig2} that the couplings $g_{ZH^0A^0}$ and $g_{ZZh^0}$ are also involved in each of the scattering processes. Therefore, these couplings need to be determined as well.
    Studying $e^-e^+\rightarrow Zh^0$ and also the process  $e^-e^+\rightarrow ZZh^0$ with a smaller cross section could let us determine the coupling $g_{ZZh^0}$ \cite{PhysRevD.94.113002} , the same is true for the process $e^-e^+\rightarrow A^0h^0$ and $e^-e^+\rightarrow ZA^0h^0$ which makes it possible to determine the coupling $g_{ZH^0A^0}$ \cite{2010arXiv1001.0473L, 2012arXiv1204.1834L}.

\section{Machinery for the numerical analysis}
\label{sec3}
    
    The scattering of all the processes are denoted as 
    \begin{eqnarray}
    e^- (k_1,\mu)\; +\;e^+ (k_2,\nu)\;\rightarrow \;  A (k_3)\; +\; B (k_4) \;+\; C (k_5)
    \label{eq:eq4}
    \end{eqnarray}
    where $k_a$ $(a=1,...,5)$ are the four-momenta of the incoming electron and the positron beam, at the right-hand side of the reaction A, B and C represent the final states defined in table \ref{tab:tab2}, respectively. The spin polarizations of the incoming particles are denoted by $\mu$ and $\nu$. The relevant Feynman diagrams which make the contribution to the scattering in SM and 2HDM are shown in figure \ref{fig1} and \ref{fig2}, respectively. 
    The vertices are calculated with the help of \textsc{FeynRules} \cite{Alloul:2013bka, Degrande:2014vpa}, and they are in good agreement with 
    the model file in \textsc{FeynArts} \cite{Kublbeck:1992mt, Hahn:2000kx}, then the diagrams as well as the amplitudes are obtained employing \textsc{FeynArts}. After, the simplification of the amplitudes, squaring the total amplitude, and integration over the phase space of the final states in a $2\rightarrow3$ scattering is accomplished using the driver program in \textsc{FormCalc} \cite{Hahn:2006qw} routines.

    The differential cross section for each of the scattering processes, which are given in table \ref{tab:tab2}, are defined as
    \begin{equation}\label{eq:partcross}
    d\sigma (s;\mu,\nu)=\frac{1}{n!} \frac{1}{\Phi(s)} \left(\frac{1}{4} \sum_{hel}{|\mathcal{M}(s;\mu,\nu)_{tot}|^2}\right) d\Phi^{(3)}
    \end{equation}
    where $\Phi(s)=\sqrt{s^2-4sm_e^2}$ is the flux factor for the incoming $e^-e^+$ beams. $\mathcal{M}$ is the total amplitude of all the tree-level diagrams for each processes. The factor $1/n!$ is due to the identical particles at the final state. The summation in equation \ref{eq:partcross} is taken over the polarization of the Z-boson if there is, and next the spin-averaging of the initial particles are employed. The three-particle phase-space of the final state is defined as
    \begin{equation}
    d\Phi^{(3)}=\delta\left(k_1+k_2-\sum_{i=3}^5k_i\right)\prod_{j=3}^5\frac{d^3k_j}{(2\pi)^3 2E_j}.
    \label{eq:}
    \end{equation}
    The computation requires a mutli-dimensional integration, and we employed Monte-Carlo integration methods. Therefore, the routines in \textsc{CUBA} \cite{Hahn:2005pf,Hahn:2016ktb} library are used. 

    The polarized cross section $\sigma(s;P_{e^+}, P_{e^-})$ for an arbitrary degree of longitudinal beam polarizations is defined as 
    \begin{equation}
    \sigma(s; P_{e^-}, P_{e^+} )= \frac{1}{4}     \left[ (1-P_{e^-})(1+P_{e^+})\sigma_{LR} + (1+P_{e^-})(1-P_{e^+})\sigma_{RL} \right],
    \label{eq14}
    \end{equation}
    where $\sigma_{LR}$ stands for the cross section where the electron beam is polarized completely left-handed ($P_{e^+} = -1$), and the positron beam is polarized completely ($P_{e^-} = +1$) right-handed. The cross sections $\sigma_{RL}$, $\sigma_{LL}$, and $\sigma_{RR}$ are defined similarly. Note that due to the nature of the scattering process, $\sigma_{LL}$ and $\sigma_{RR}$ are small to make an impact so we safely neglected these contributions in equation \ref{eq14}. 
 
\section{The cross section distributions and discussion}
\label{sec4}

    In the computation, the results are presented for the following constants. The SM parameters are taken from reference \cite{Eidelman:2004wy} where $m_e=0.51099\text{ MeV}$, $m_Z=91.1876\gev$, $s_w=0.222897$, and $\alpha=1/127.944$ are given. The mass of the SM Higgs boson is $m_H=125.09\gev$ \cite{Aad:2012tfa, Chatrchyan:2012xdj, Khachatryan:2014jba}. The other free parameters in the 2HDM are already introduced in section \ref{sec2}. 
    
\begin{figure}[htbp]
    \centering 
    \includegraphics[width=0.40\textwidth]{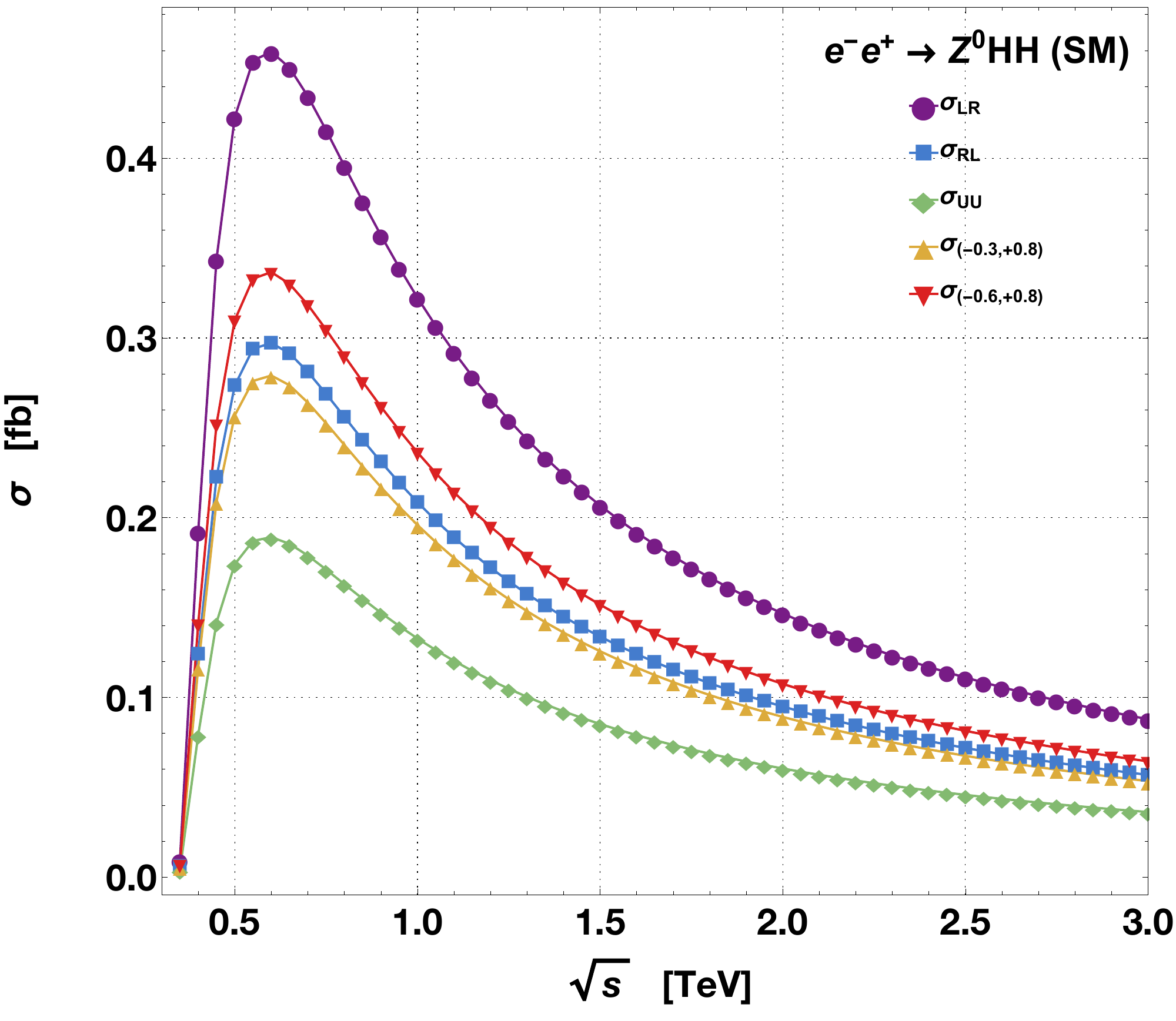}
    \hspace{0.5cm}
    \includegraphics[width=0.42\textwidth]{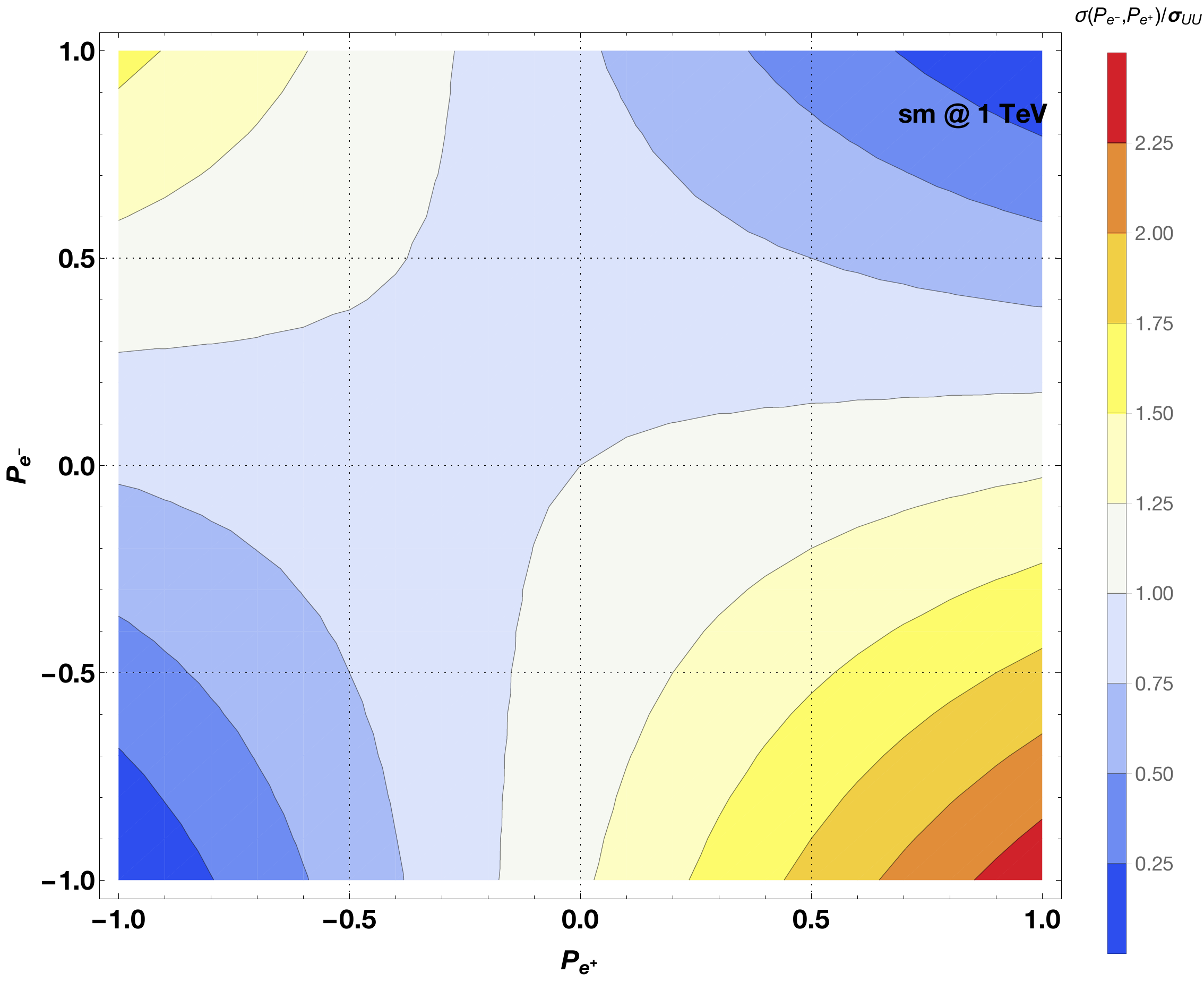}
    \caption{\label{fig3} 
    (left): The distributions of the cross section for various polarizations of the incoming $e^-e^+$ beams. 
    (right): The ratio of the cross sections ($\sigma(P_{e^-},P_{e^+})/\sigma_{UU}$) as a function of the polarization of the incoming electron/positron beams.
    }
    \end{figure}

    The cross section of the prominent channel $e^-e^+\rightarrow Z^0HH$ (double Higgs-strahlung) in SM is presented in figure \ref{fig3} (left). The unpolarized cross section is around $0.174 \fb$ at $\sqrt{s}=0.5\tev$, and it rises to $0.189 \fb$ at $\sqrt{s}=0.6\tev$, then decreases slowly for higher energies. It is also seen that the left-handed polarized electron and the right-handed polarized positron enhance the cross section up to $0.459\fb$. Additionally, the distributions for two polarization cases ($\sigma(-0.3,+0.8)$) and ($\sigma(-0.6,+0.8)$) are given in figure \ref{fig3} (left). Besides of that, on the right-hand side of the figure \ref{fig3}, all the possible polarization configurations for the incoming beams are scanned, and the cross section for double Higgs-strahlung is computed. The ratio $\sigma(P_{e^-},P_{e^+})/\sigma_{UU}$ is plotted as a function of ($P_{e^-},P_{e^+}$) using the equation \ref{eq14}. It clearly shows that, the left-handed electron beam ($P_{e^-}=-1$) and right-handed positron beam ($P_{e^+}=+1$) maximize the cross section. The enhancement in the cross section is raised up to a factor of 2.25 at the right bottom corner. However, it is dropped significantly at the left bottom and the top right corners. 

    In this study, the exact alignment limit is taken for the 2HDM. As a result, the triple Higgs coupling $g_{h^0h^0h^0}$ gets the same form with the coupling $g_{HHH}$ in the SM. Therefore, the process $e^-e^+\rightarrow Z^0h^0h^0$ in 2HDM has the same distribution given in figure \ref{fig3} (left). Hence, an additional figure with the same distribution is not plotted for this process. It is clear that a future lepton collider which has a c.m. energy of $1\tev$ could easily probe the Higgs self-coupling $g_{h^0h^0h^0}$. As it is intended by the exact alignment limit, $h^0$ has the same couplings and the same production cross section with the SM Higgs boson regarding the process ${Z^0h^0h^0}$. 
    
\begin{figure}[htbp]
    \centering
    \includegraphics[width=0.32\linewidth]{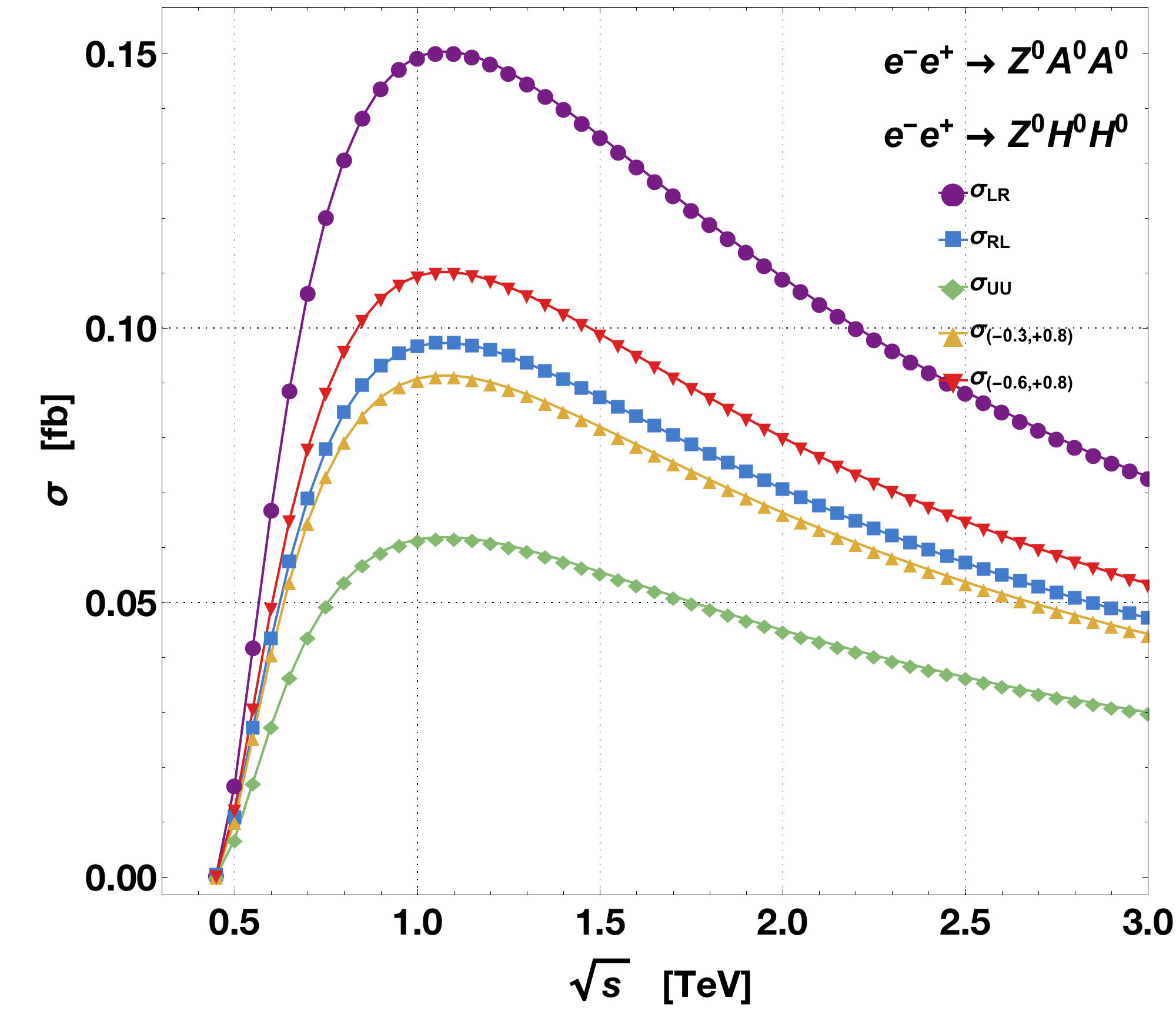}
    \includegraphics[width=0.32\linewidth]{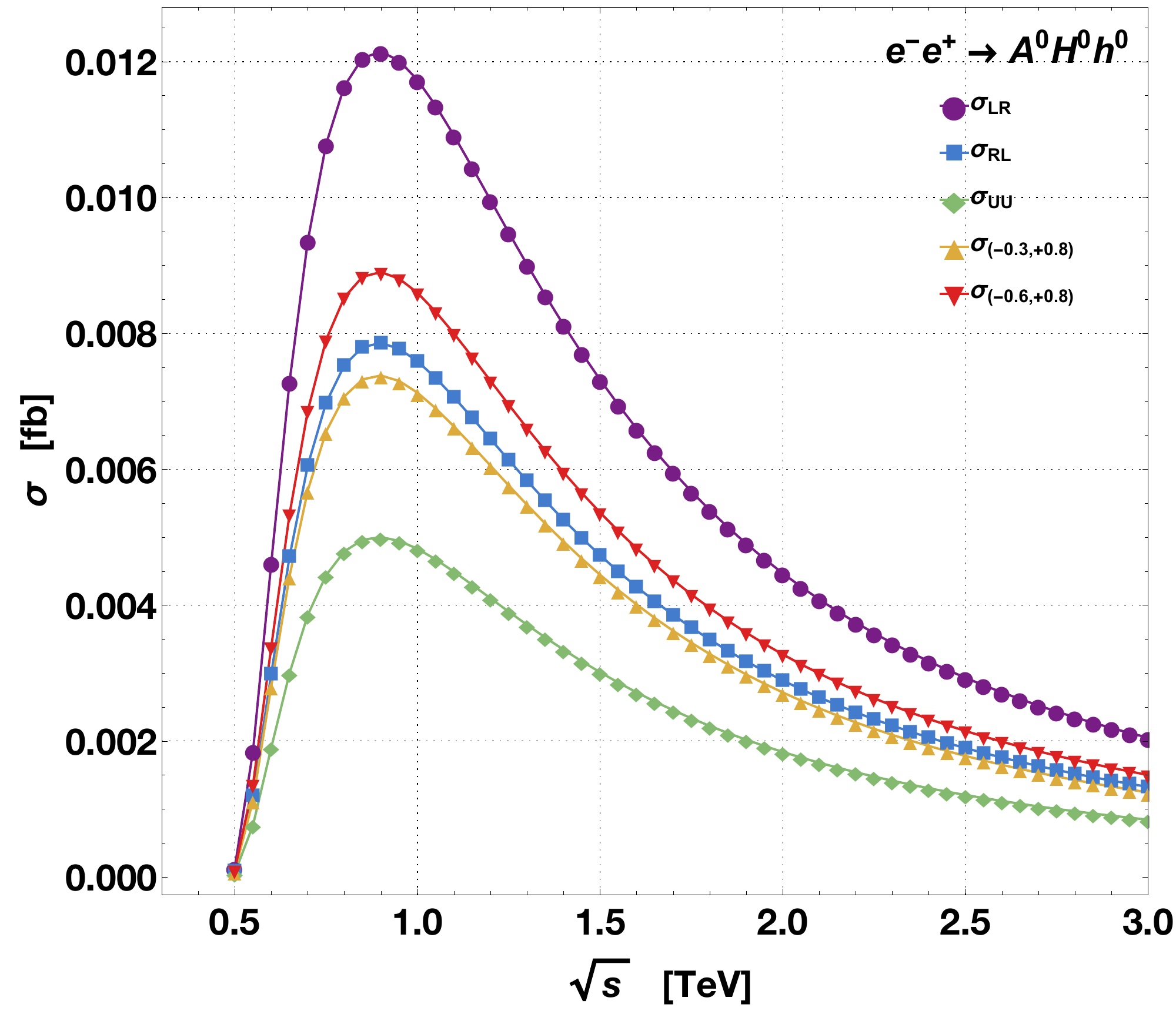}
    \includegraphics[width=0.32\linewidth]{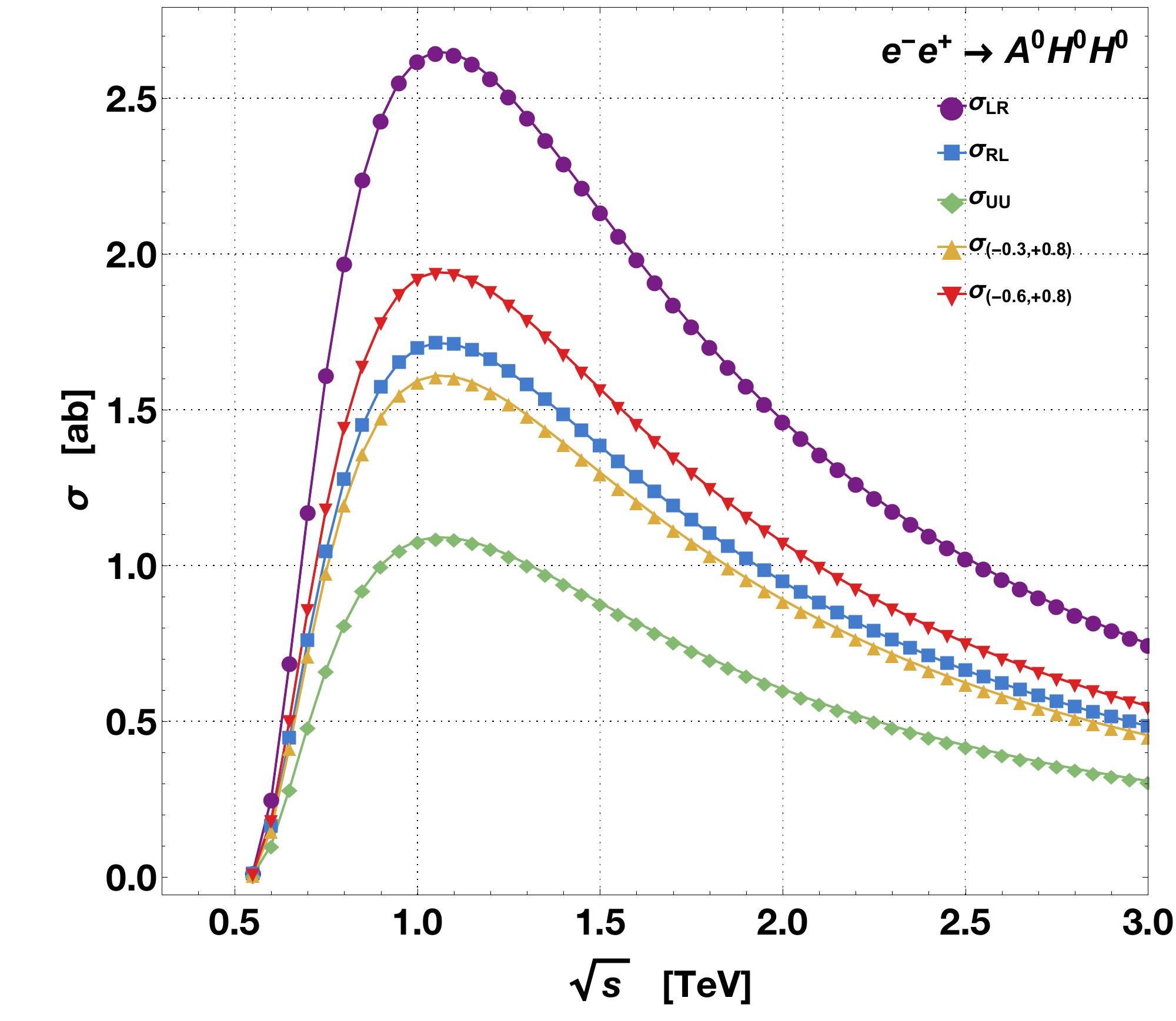}
    \caption{\label{fig4}  The distributions of the cross section for various polarizations of the incoming $e^-e^+$ beams.
    The mass of the extra Higgs states is set as $m_{H}=175\gev$, and $\tb=10$ is assumed.
    (left): The cross section distributions for $e^-e^+\rightarrow Z^0A^0A^0$ and $e^-e^+\rightarrow Z^0H^0H^0$ are given.
    (center): The distributions are for $e^-e^+\rightarrow A^0H^0h^0$.
    (right): The process $e^-e^+\rightarrow A^0H^0H^0$ is plotted.
    }
    \end{figure}

    Considering the parameter space, the couplings $g_{h^0H^0H^0}$ and $g_{h^0A^0A^0}$, as well as the masses of $m_{H^0}$ and $m_{A^0}$ are identical. In addition to that, the topology of Feynman diagrams which take place in $e^-e^+\rightarrow Z^0A^0A^0$ and $e^-e^+\rightarrow Z^0H^0H^0$ scattering processes are the same. Therefore, the distribution of the cross section becomes identical, and they are plotted for various polarization cases in figure \ref{fig4} (left). Consequently, the distributions given in figure \ref{fig4} (left) hold for these two processes. The unpolarized cross section reaches $\sigma_{UU}(e^-e^+\rightarrow Z^0A^0A^0/ Z^0H^0H^0) \sim 0.062\fb$ around $\sqrt{s}=1\tev$. Then, it falls slowly at higher energies. Consequently, these two processes will be the next ones to study in the future lepton colliders, and they could be used to extract the couplings $g_{h^0A^0A^0}$ and $g_{h^0H^0H^0}$. 
    Moving to the next process, the distributions are given in figure \ref{fig4} (center) for $e^-e^+\rightarrow A^0H^0h^0$. The cross section is $\sigma_{UU}\sim0.005\fb$ at $\sqrt{s}=1\tev$. This process could let us to extract the coupling $g_{H^0A^0A^0}$, but the cross section is quite small. 

    Finally, the cross section is calculated for $e^-e^+\rightarrow A^0H^0H^0$ and plotted in figure \ref{fig4} (right). Compared to the other processes $e^-e^+\rightarrow A^0H^0H^0$ has the smallest cross section. The cross section for the unpolarized incoming beams is $\sigma_{UU}\sim 1.1\ab$ at $\sqrt{s}=1\tev$, and it drops rapidly at higher energies. It should be underlined that there are two couplings involved in the scattering which are $g_{H^0A^0A^0}$ and $g_{H^0H^0H^0}$, and both of them are a function of $\Lambda_7$. The polarization of the incoming beam has the potential to enhance the cross section which could improve the number of events to be detected at the future lepton colliders. Considering the total luminosity which will be gathered, it will be hard to measure or extract the coupling $g_{H^0H^0H^0}$. On the other hand, as it is mentioned before that the couplings $g_{ZZh^0}$ and $g_{ZH^0A^0}$ are necessary to obtain the Higgs self-couplings fully. Indeed, the following scattering processes $e^-e^+\rightarrow ZZh^0$ and $e^-e^+\rightarrow ZA^0h^0$ have a sole function of determining the couplings $g_{ZZh^0}$ and $g_{ZH^0A^0}$, respectively. These process have a cross section of $\sigma_{UU}(e^-e^+\rightarrow ZZh^0)=0.288\fb$ and $\sigma_{UU}(e^-e^+\rightarrow ZA^0h^0)=0.098\fb$ at $\sqrt{s}=1\tev$.

\begin{figure}[htbp]
    \centering 
    \includegraphics[width=0.40\textwidth]{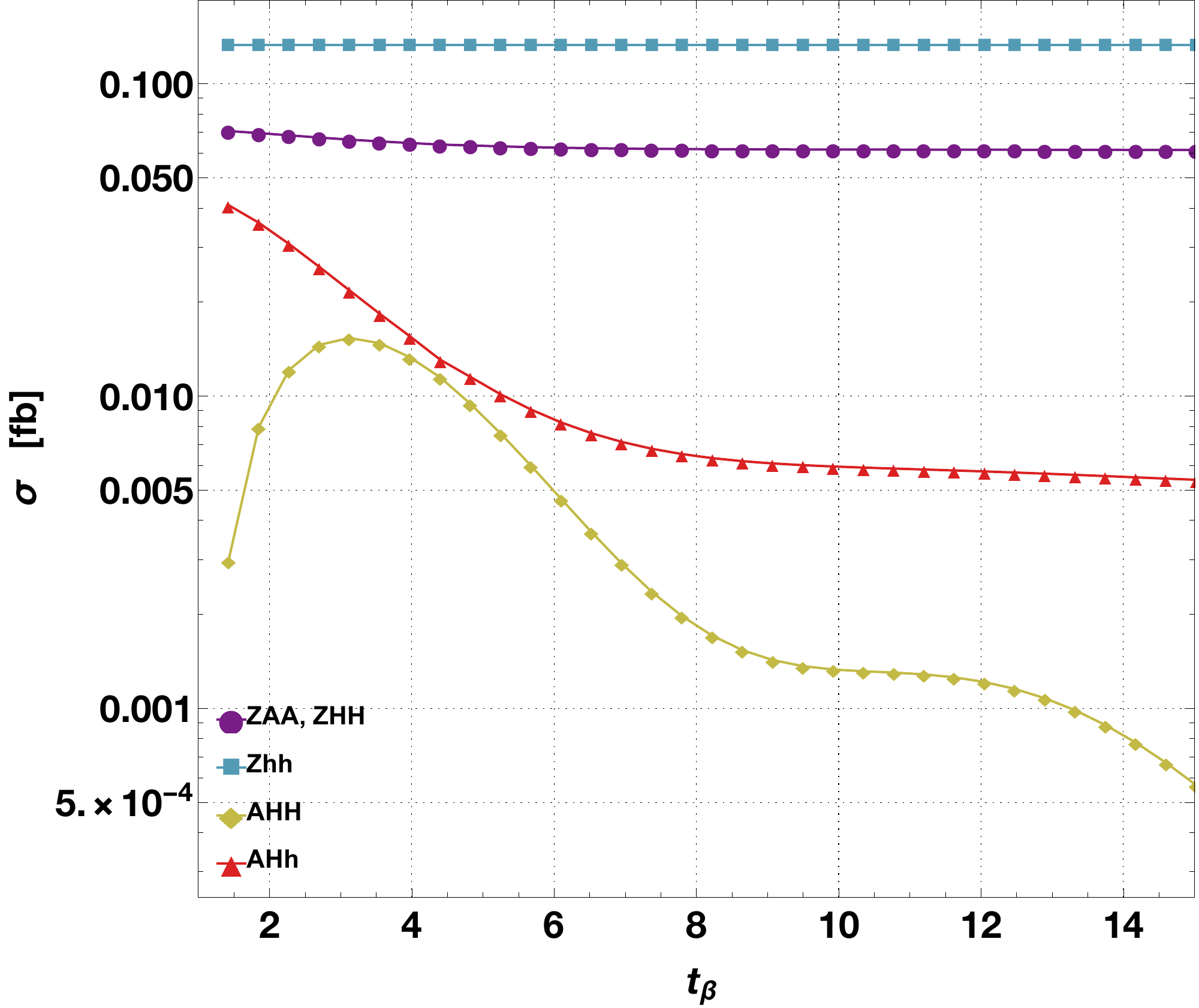}
    \hspace{0.5cm}
    \includegraphics[width=0.40\textwidth]{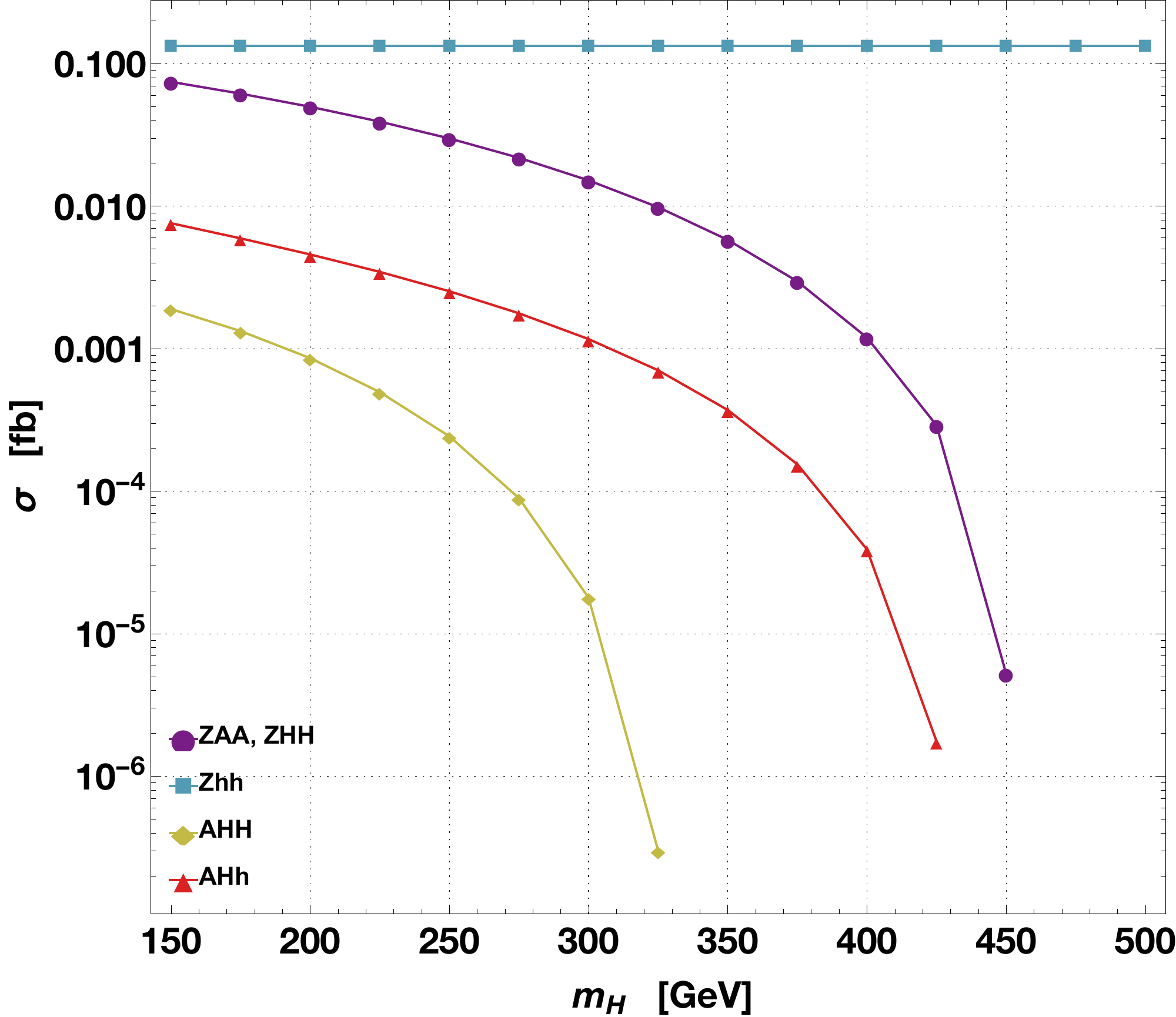}
    \caption{\label{fig5} 
    The distributions of the cross sections for all the processes as a function of $\tb$ (left) and $m_H$ (right) at $\sqrt{s}=1\tev$. The mass of the extra Higgs states is set to $m_{H}=175\gev$ (left) and $\tb=10$ (right), respectively.
    }
    \end{figure}

     An analysis is also carried out to test the $\tb$ dependence, and it is given in figure \ref{fig5} (left) at $\sqrt{s}=1\tev$. It is seen that the cross section is flat for the process $e^- e^+ \rightarrow Z h^0h^0$, that is already expected because the coupling $g_{h^0h^0h^0}$ is the same as the SM one, and it does not change with the $\tb$. In the exact alignment limit ($\sba=1$), the production of $ZA^0A^0$ and $ZH^0H^0$ have identical distributions, since both processes are a function of the same factor ($\Lambda_3$) that dependence is foreseen. Next, the production rate of $A^0H^0h^0$ is at the maximum at low $\tb$, then it falls at higher $\tb$ values and reaches to saturation for $\tb>8$. The last distribution is the production of $A^0H^0H^0$. Since two couplings are involved in this process, the $\tb$ dependence is similar at high $\tb$ values with the $A^0H^0h^0$ final state. While the cross section is rising with decreasing values of $\tb$, it falls again for $\tb<3$.
     Finally, the cross section gets declined for all the processes investigated in table \ref{tab:tab2} at increasing $m_{H}$ values given in figure \ref{fig5} (right). That is anticipated because the mass of all the extra Higgs states is increased, and the phase space becomes narrowed for the particles at the final state. There is one exception which is the production of $Zh^0h^0$, that process does not depend on the $m_H$ mass, and it is flat for all $m_H$ values. Overall, the production cross section of the other processes increases for small $m_H$ values.
          

\section{Identifying the process at the detector}
\label{sec5}

    In this section, the decay channels of each of the Higgs bosons are discussed, and possible collider signatures for measuring each of the processes are examined. The possible background channels, the number of events expected in benchmark luminosities, and challenges are indicated for the detection of the processes in a collider. 

\subsection{The decays of the neutral Higgs bosons}

\begin{figure}[htbp]
    \centering 
    \includegraphics[width=0.325\textwidth]{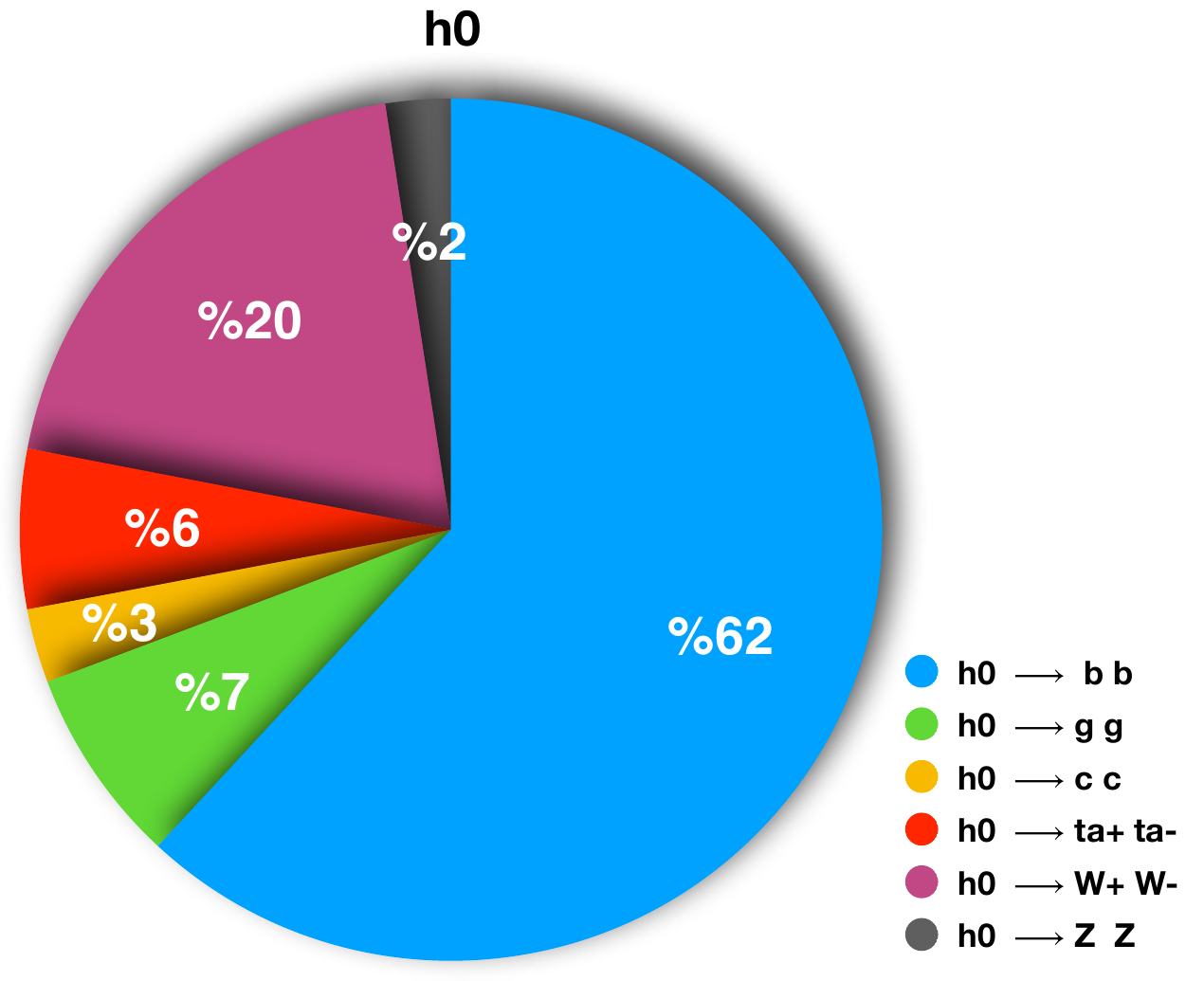}
    \includegraphics[width=0.325\textwidth]{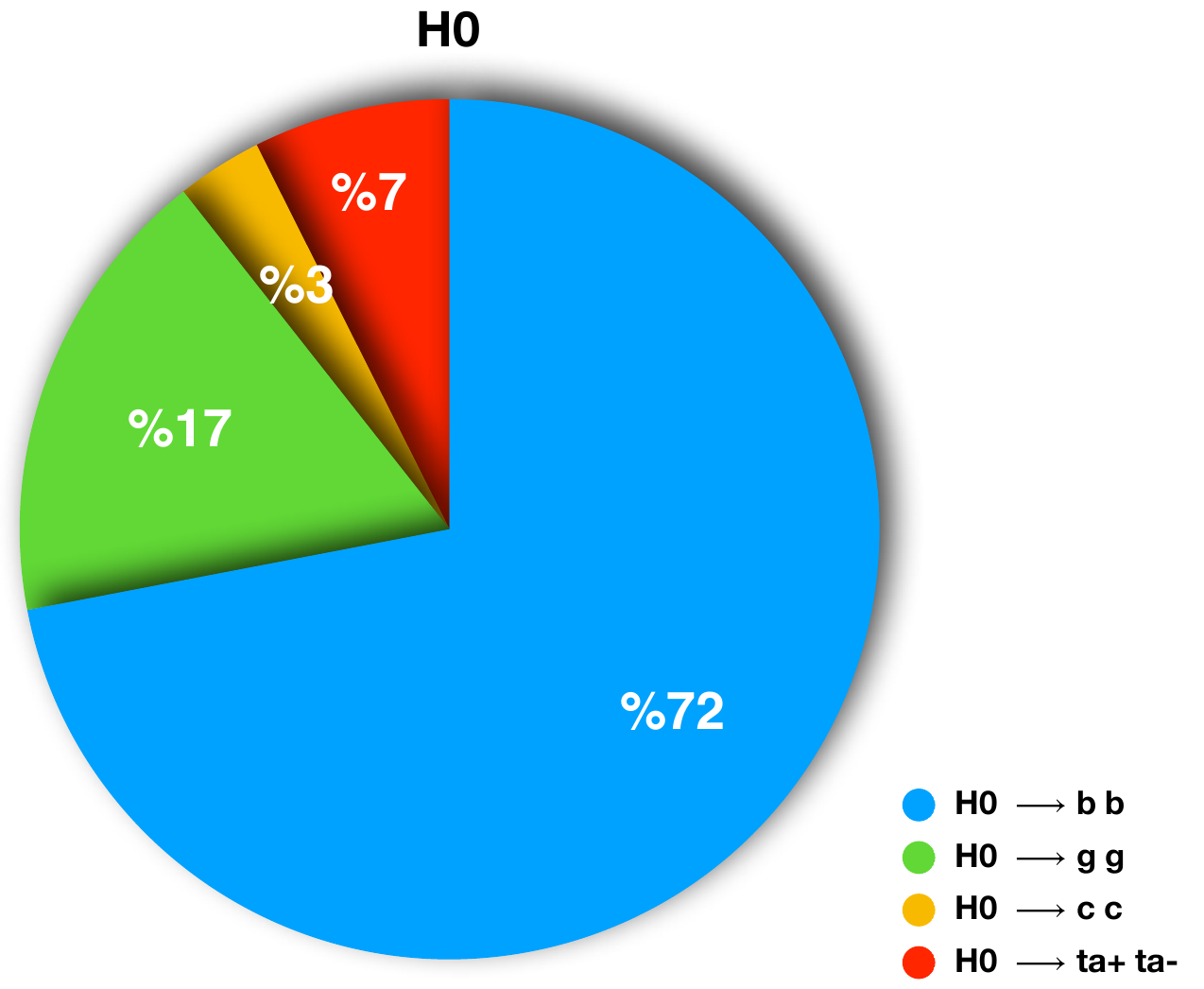}
    \includegraphics[width=0.325\textwidth]{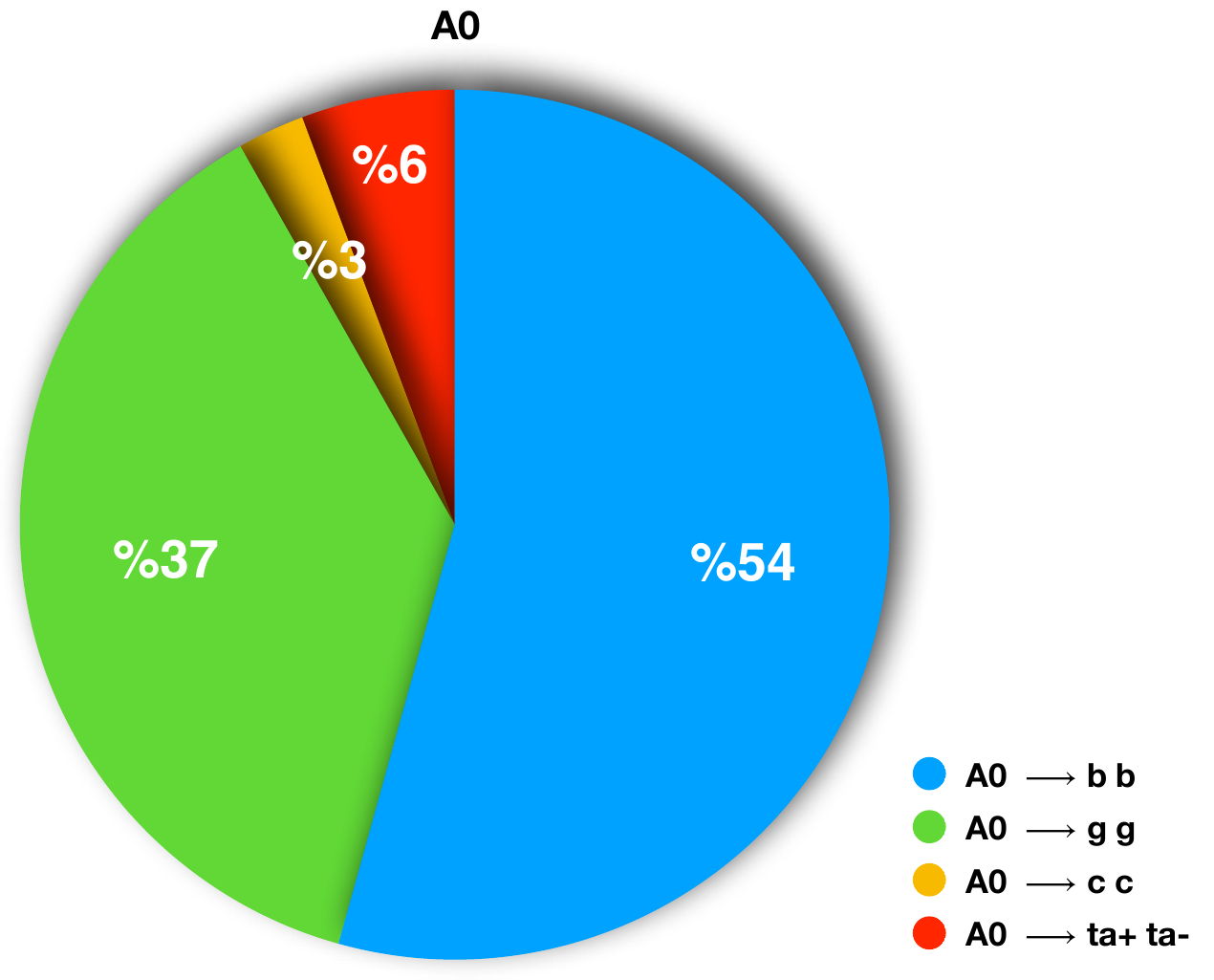}
    \caption{\label{fig5a} 
    The branching ratios of all the neutral Higgs bosons for the point defined in table \ref{tab1}.}
    \end{figure}

     The decay widths and the branching ratios of all the Higgs bosons are calculated for the region defined in table \ref{tab1} using the \textsc{2HDMC}. In figure \ref{fig5a}, the branching ratios of the Higgs particles are given as a pie chart for the neutral Higgs bosons. The decay width of each of the Higgs bosons depends on the relevant vertices and the masses of the particles involved, but the branching ratios are stable. Besides, these decay channels for each of the Higgs bosons are the same in varying the Higgs mass $m_{H}$. It can be seen in figure \ref{fig5a}, the dominant decay channel for all the neutral Higgs bosons is through $b\bar{b}$-pair, and $\mathcal{BR}(h^0/H^0/A^0\rightarrow b\bar{b})\approx (62,72,54)\%$. Then, the second and the third dominant ones are gluon and $c\bar{c}$ pairs, respectively. It is logical to say that, the dominant pattern for each of the neutral Higgses in the detector is the di-jet due to the b-quark, the gluon, and the c-quark pairs. The fourth decay channel that is common for all of them is the $h^0/H^0/A^0\rightarrow\tau\bar{\tau}$, but the branching ratio is low compared to the di-jet signal. The lightest Higgs, which resembles the SM Higgs boson, has other decay channels through vector boson pairs that are considerably large compared to the other neutral Higgs bosons. Even though these decay channels could be considered as advantageous, they could be cumbersome due to the extra Z-boson at the final state and the leptonic decays of the W-boson. Therefore, the hadronic decay channels of $h^0$ boson promise more in the extraction of its pattern. 
    
\subsection{Identification of the processes and possible background channels}

    Since the decay channels for each of the particles at the final state are defined previously, the pattern for each of the channels that are expected at the detector could be obtained easily. The Z-boson decays through three main channels with the following branching ratios: hadronic $\mathcal{BR}(Z^0\rightarrow q\bar{q})\sim 0.70$, leptonic $\mathcal{BR}(Z^0\rightarrow l\bar{l})\sim 0.10$, and invisible ($\sim 0.20$) \cite{Eidelman:2004wy}. An ideal case for the channels where the Z-boson is presented would be letting all the neutral Higgs bosons decay through $b\bar{b}$ pair + the hadronic decay of the $Z^0$ boson. Thus, there will be 4 b-quark initiated (b-tagged) jets + 2 light jets (coming from the Z-boson decay) at the final state. 
    Additionally, it is possible to trigger the events with $Z^0\rightarrow l\bar{l}$ instead of hadronic decay, and there will be two leptons with the opposite sign in the final state + 4 b-tagged jets. Unfortunately, the leptonic branching ratio is small compared to the hadronic decays, and there could be as much as seven times fewer events accumulated in the detector. Since b-tagging algorithms let to distinguish the flavor of jets, it would be useful to explore various possible final states. Among all the possible patterns at the detector, the six of them which are distinguishable are explored, and the percentage of the events are given in table \ref{tab3}. It is seen that the full hadronic decay of all the particles at the final state gives the biggest fraction of events for each of the processes. However, the full hadronic final state of the $Z^0h^0h^0$ has the lowest percentage because it has more options to decay such as $W^+W^-/Z^0Z^0$ (figure \ref{fig5a} (left)). There are some detector level studies explored these channels \cite{Asner:2013psa, Tian:2013qmi}. It should be noted that a full detector simulation of  $e^-e^+\rightarrow Z^0HH$ in SM was performed in the following references \cite{GutierrezRodriguez:2008nk, Battaglia:2001nn,  dEnterria:2016sca, Castanier:2001sf}.

    \begin{table}[htp]
    \caption{The percentage of the events with different decay channels are given. Since there is no Z-boson in the last two processes, the stared numbers refers to $\tau\bar{\tau}$ decay channel in $l\bar{l}$.\label{tab3}
    }
    \centering
    \begin{tabular}{ l | l || c | c | c | c | c }
    \hline  & Detector patterns
                              & $ Z^0A^0A^0$
                                  & $ Z^0H^0H^0$ 
                                      &  $ Z^0h^0h^0$   
                                          & $A^0H^0h^0$ 
                                               & $ A^0H^0H^0$
                                                \\
    \hline\hline               
   1 & 4 b-quark jets + 2 jets              &    20.59    &    36.24    & 26.60    &    28.54    &    36.74    \\ \hline
   2 & 4 b-quark jets + $l\bar{l}$          &    2.94     &    5.18     & 3.80     &    7.33*    &    8.73*    \\ \hline
   3 & 2 b-quark jets + $l\bar{l}$ + 2 jets &    4.33     &    2.96     & 1.25     &    5.75*    &    7.60*    \\ \hline
   4 & 2 b-quark jets + 4 jets              &    30.32    &    20.74    & 8.72     &    9.10*    &    14.13*   \\ \hline
   5 & 4 jets + $l\bar{l}$                  &    8.87     &    8.56     & 5.15     &    13.98*   &    17.79*   \\ \hline
   6 & 6 jets                               &    62.08    &    59.94    & 36.03    &    62.53    &    80.64    \\ \hline
    \end{tabular}
    \end{table}%
    
    If it is assumed that the ILC project could obtain a total integrated luminosity of $1 \iab$ (or $3 \iab$ in high lumi phase) \cite{Fujii:2017ekh} in its lifetime, then the number of events expected for each of the processes at $\sqrt{s}=1\tev$ is given in table \ref{tab4} where the extra Higgs masses $m_{H}=175\gev$ and $\tb=10$ are set. The number of events expected for each of the patterns could be calculated easily using table \ref{tab3} and table \ref{tab4}. Accordingly, if 6 jets final state is considered with $3 \iab$ total luminosity, one expects a total of about 115 (111) events in $Z^0A^0A^0$ ($ Z^0H^0H^0$) final state. That is without taking into account the experimental acceptance cuts and various efficiencies. The same calculation yields about 144 events for $Z^0h^0h^0$ with $\mathcal{L}=3 \iab$. Unfortunately, the last two scattering processes $A^0H^0h^0$ and $A^0H^0H^0$ yield $\lesssim 10$ and $\lesssim 3$ events with $\mathcal{L}=3 \iab$, respectively. The polarization of the incoming beams could increase the number of events up to 2.25 times at most (figure \ref{fig3} (right)). However, that still could not be enough to measure these two processes. Eventually, exploring the full hadronic final state for each process gives more events in the detector, and it could be the best chance to identify these processes. However, all the possible background channels need to be considered as well.

    \begin{table}[htp]
    \caption{The expected number of events for two benchmark luminosities at $\sqrt{s}=1\tev$, {where $m_{H}=175\gev$ and $\tb=10$ are set.}\label{tab4}
    }
    \centering
    \begin{tabular}{ l || c | c | c | c | c }
    \hline  
                              & $ Z^0A^0A^0$ 
                                  & $ Z^0H^0H^0$
                                      & $ Z^0h^0h^0$    
                                          & $A^0H^0h^0$ 
                                               & $ A^0H^0H^0$
                                                \\
    \hline\hline                
    $\mathcal{L}=1  \iab$   &  62    & 62        & 133 & 5 &  1.1         \\ \hline
    $\mathcal{L}=3  \iab$   &  186    & 186    & 399 & 15 &  3.3     \\
    \hline 
    \end{tabular}
    \end{table}%

    The production rate of each of the processes in $e^+e^-$ collider is small. Considering the weakness of the signals, it could be asked whether these processes could be extracted from the SM background. In table \ref{tab3}, it is seen that all the neutral Higgses decay through the b-quark pair, and it indicates that the b-quark identification is vital in the reconstruction of each of the processes and also in some of the production channels. Also, due to the parton branching of quarks/gluons and the missefficiency of b-tagging algorithms, the number of jets or b-tagged jets are not fixed in the final state. Therefore, observing some of these patterns at the detector and making measurements of the Higgs self-couplings will be challenging.

    There are also many background channels which could hide the processes we are interested in. Some of the background channels most relevant and expected to shadow the processes are as follows: $\epem Z\bb\bb$, $\epem Z\bb\cc$, $\epem Z\cc\cc$, and $\epem ZZ\rightarrow \bb\bb$ in SM. Therefore, reconstructing the Higgs masses in each event could be useful. If the b-quark pairs do not come from the neutral Higgses, they will fall out of the Higgs mass range, and these events could be excluded. If the b-tagging efficiency is taken around 80\% or higher, and requiring $H_iH_j\rightarrow \bb\bb$ + hadronic decay of Z-boson offers a distinct pattern at the detector which is 4 b-tagged jets + 2 light jets. That pattern has a significant fraction of the events, and because of b-tagged jet requirement, it could be used to eliminate the big fraction of the background channels. 
    Moreover, if the top-quark (Z-boson and neutral Higgs bosons) is demanded to decay into semi-leptonic (hadronic) final states, top-quark involved processes will also contribute to the main background: 
        $\epem\tti\cc$,
        $\epem\tti\bb$, 
        $\epem\tti Z( \rightarrow \bb)$, 
        $\epem\tti H (\rightarrow \bb)$, 
        $\epem t\bar{t}ZH$, 
        $\epem t\bar{t}ZZ$,
        $\epem\tti$, and  
        $\epem\tti j$ where the top-quarks and the Higgs boson decaying through the $\bb$ pair or the light jets could mimic some of the final states of the processes given in table \ref{tab3}. Requiring the top-quark reconstruction and kinematical cut on these events could eliminate a large fraction of them. Light jets associated with a various number of vector bosons could also be considered as a background. In any case, a Monte Carlo simulation of each of the processes with different decay channels is required to estimate the trigger efficiency and the acceptance of the detector. Thus, a realistic estimation of the potential of the future lepton colliders could be obtained.

\section{Summary and conclusion}
\label{sec6}

    In this study, the production rate of various processes is carried out in a $e^+e^-$ collider. These processes are selected for extracting the triple Higgs self-couplings in the 2HDM. The model is examined considering the new experimental constraints on the charged Higgs boson. These constraints favor the exact alignment limit where $\sba=1$, and consequently $h^0$ becomes indistinguishable from the SM Higgs boson. There are in total of eight possible Higgs self-couplings, and two of them includes the charged Higgs states which are not within the scope of this study. One of the rest vanishes when $\sba=1$, thus, only five of them survive. The involvement of the Higgs self-couplings for each of the processes are given in table \ref{tab:tab2}. 
    As we deliberately picked the $\sba=1$ limit, the scattering process $e^-e^+\rightarrow Z^0h^0h^0$ helps to extract the prominent coupling $g_{h^0h^0h^0}$ just like in the SM. The next task, in the extraction of the triple self-couplings, would be studying $Z^0A^0A^0$ and $Z^0H^0H^0$ final states, then the couplings $g_{h^0A^0A^0}$ and $g_{h^0H^0H^0}$ could be determined, respectively. Besides, these final states have a modest cross section, and the plan in the colliders would be studying them after $Zh^0h^0$.    
    The next process $e^-e^+\rightarrow A^0H^0h^0$ lets us access the coupling $g_{H^0A^0A^0}$. However, the cross section is small, and it might not be possible to collect enough events.
    Finally, the process $e^-e^+\rightarrow A^0H^0H^0$ makes it possible to determine $g_{H^0H^0H^0}$ with the help of the coupling $g_{H^0A^0A^0}$ if it ever could be obtained in the previous step. However, considering the previous process is hard to observe, measurement of the $g_{H^0H^0H^0}$ coupling is also a challenge using the processes investigated in this paper. On the other hand, various polarization scenarios of the incoming beams have a potential to increase the cross section. Indeed, the cross section is enhanced up to a factor of 1.8 for $P{(e^-e^+)}=(-0.60,+0.80)$, and the polarization has the same effects on all the processes. 
            
    The decay channels of all the neutral Higgs bosons and possible patterns of each of the processes are also investigated. The analysis shows that in this particular choice of the parameter space all the neutral Higgs bosons have similar decay channels. They mainly decay through $\bb$ pair, light quark pair, gluon pair, and with a small fraction to $\tau\bar{\tau}$ pair. Therefore, if one chooses to study the hadronic decays of all the scattering processes, then the highest fraction of events could be obtained in the detector. It is concluded that the jet-finding algorithms will determine whether these processes with the given patterns could be observed due to the challenges in the high jet multiplicity environment in the detector. The jet reconstruction and better efficiency of b-quark initiated jets for some of the channels are essential to reconstruct the processes. A better assessment of the observability of each of the processes requires Monte Carlo simulation of the signal and all the possible background processes, that is beyond the scope of this paper.
  
    Among the proposals of all the future lepton colliders, the ILC with c.m. range up to 1 $\tev$ has the biggest potential regarding all the cross section distributions. However, the FCC-ee with a $\sqrt{s}=0.5\tev$ could still make measurements and compete for the couplings $g_{h^0h^0h^0}$, $g_{h^0A^0A^0}$ and $g_{h^0H^0H^0}$. Unfortunately, the proposed CEPC do not have enough c.m. energy to explore the Higgs self-couplings, even the process $e^-e^+\rightarrow Z^0HH$ in the SM. 

    The last missing piece of the SM (Higgs boson) is exposed by the LHC. However, there is no any clue to the new physics. The 2HDM is one simple extension of the SM with full of predictions. This study shows the potential of exploring the triple Higgs self-couplings in the 2HDM in the future lepton colliders. Measuring these couplings will let us confirm the shape of the Higgs potential. However, obtaining the triple Higgs self-couplings is not enough for determining the shape of the Higgs potential. The complete reconstruction could be achieved by measuring the quartic Higgs self-couplings as well.
    
\acknowledgments
    The computation presented in this paper was partially performed at TUBITAK ULAKBIM, High Performance and Grid Computing Center (\textsc{TRUBA} resources). Part of is also accomplished at the computing resource of \textsc{FENCLUSTER} (Faculty of Science, Ege University). Ege University supports this work, project number 17-FEN-054.

\bibliographystyle{JHEP}

\bibliography{template-8s_revtex}

\end{document}